\documentclass[twocolumn,twocolappendix]{aastex631}

\usepackage[toc,titletoc]{appendix}
\usepackage {appendix}
\usepackage{threeparttable}
\usepackage{xcolor,graphicx,xspace,color,longtable,enumitem}
\usepackage{amssymb,amsmath,shadow,bezier,curves,rotating}
\usepackage{graphicx,chngcntr}
\graphicspath{ {./images/}}

\newcommand {\h}  {$h^{-1}\,$Mpc}

\newcommand {\hmm} {$h^{-2} \  M_{\odot}$}
\newcommand {\m}  {M$_{\odot}$}

\newcommand {\M} {\mathcal{M}}
\newcommand {\N} {\mathcal{N}}

\begin{document}

\title[Size-Mass Relation]{From the Densest Clusters to the Emptiest Voids: No Evidence For Environmental Effects on the Galaxy Size–Mass Relation at Low Redshift}

\author[0000-0003-3595-7147]{Mohamed H. Abdullah}
\affiliation{Department of Physics, University of California Merced, 5200 North Lake Road, Merced, CA 95343, USA}
\affiliation{Department of Astronomy, National Research Institute of Astronomy and Geophysics, Cairo, 11421, Egypt}

\author[0009-0003-6781-1895]{Nouran E. Abdelhamid}
\affiliation{Department of Astronomy, National Research Institute of Astronomy and Geophysics, Cairo, 11421, Egypt}

\author[0000-0003-2716-8332]{Rasha M. Samir}
\affiliation{Department of Astronomy, National Research Institute of Astronomy and Geophysics, Cairo, 11421, Egypt}

\author[0000-0002-6572-7089]{Gillian Wilson}
\affiliation{Department of Physics, University of California Merced, 5200 North Lake Road, Merced, CA 95343, USA}

\begin{abstract}
We present a comprehensive study of the galaxy size–stellar mass relation (SMR) at low redshift ($z \leq 0.125$), using a large spectroscopic sample from the SDSS-DR13 survey. Our goal is to investigate how environment affects galaxy structural properties across multiple spatial scales. Galaxies are classified by specific star formation rate, optical color and bulge-to-total light ratio, allowing us to disentangle environmental effects from intrinsic galaxy properties.
We examine the SMR in three contexts: (1) comparing galaxy sizes in two extreme environments, dense clusters versus cosmic voids; (2) analyzing cluster galaxies across a range of cluster masses; and (3) studying member galaxies located in different cluster regions, from the core to the infall zone.
In all three cases, we find no significant dependence of the SMR on environment at fixed stellar mass and galaxy type. Cluster and void galaxies follow consistent SMR trends, and no measurable variation is observed with cluster mass or cluster-centric distance.
We also confirm that early-type galaxies exhibit steeper SMR slopes than late types. Notably, this consistent lack of environmental dependence on the SMR persists even when accounting for the differing galaxy number densities in voids, supporting the universality of this SMR scaling relation across diverse environments.
\end{abstract}


\section{Introduction} \label{sec:intro}
Scaling relations between galaxy properties are fundamental tools for understanding the physical processes that shape galaxy formation and evolution \citep{Shen03,Matharu19,Chan21}. Among them, the size--stellar mass relation (SMR) links a galaxy’s physical size to its stellar mass \citep{Poggianti13,Demers2019}. The SMR captures key aspects of galaxy growth and structural transformation, showing distinctly different behaviors for star-forming and quiescent systems \citep{vanderwel2014,vanDokkum15}. It is also widely used to quantify how the structural properties of galaxy populations evolve  with redshift \citep{trujillo2004,Matharu19,Noordeh21}.

Galaxy sizes and stellar masses are shaped by a combination of internal and external processes, including gas accretion, star formation, mergers, and environmental effects \citep{Balogh04,Saracco2017}. In dense environments such as clusters, galaxies experience mechanisms like ram-pressure stripping, harassment, and tidal interactions that can quench star formation and transform morphology \citep{Dressler1984,Gnedin2003,Old20,McNab21}. These effects are especially pronounced in cluster cores, where red, quiescent ellipticals dominate, in contrast to the star-forming disk galaxies found more frequently in lower-density environments \citep{gun1972,scharf2005}.

Despite many studies exploring environmental effects on the SMR \citep{allen2017,mowla2019}, the literature presents conflicting results. Some find no clear environmental dependence, arguing that the SMR is primarily governed by intrinsic galaxy properties \citep{Weinmann2009,Huertas-Company2013a}. Others report significant environmental trends, particularly in dense regions \citep{Valentinuzzi2010,Chamba24}. These discrepancies likely stem from differences in sample selection, statistical limitations (particularly at high redshifts) and systematic uncertainties in measuring galaxy properties. For instance, variations in size and stellar mass estimates often arise from differences in light profile assumptions, the modeling of structural parameters such as Sérsic index and effective radius, and limitations in image resolution, including pixel scale, instrument quality and redshift dependent effects \citep{Shen03,Lange15}. Environmental definitions also vary significantly across studies. Some adopt statistical density thresholds \citep{Baldry06,Bolzonella10}, while others use galaxy group or cluster catalogs that offer more reliable membership assignments \citep{Yang2007}. Although statistical methods are frequently employed to distinguish high and low-density environments, they are often susceptible to projection effects and uncertainties in estimating local densities \citep{Cooper06,Muldrew12}. These methodological and systematic differences make it difficult to compare results across studies and may explain the observed inconsistencies in how environment affects the SMR.


To address these challenges, it is essential to adopt precise methods for galaxy membership determination, environmental classification and the accurate measurement of galaxy sizes and stellar masses. In this study, we examine the environmental dependence of the SMR using a sample drawn from the spectroscopic \(\mathtt{GalWCat19}\)\footnote{\url{http://cdsarc.u-strasbg.fr/viz-bin/cat/J/ApJS/246/2}} galaxy cluster catalog \citep{Abdullah20a}, constructed from SDSS DR13 data \citep{Albareti17}. The catalog employs the GalWeight technique \citep{Abdullah18}, which accounts for redshift-space distortions and avoids reliance on empirical assumptions or iterative procedures, enabling robust and accurate identification of cluster members. Furthermore, the low redshift nature of \(\mathtt{GalWCat19}\) (\(z \lesssim 0.2\)) minimizes the need to correct for evolutionary effects, making it well suited for our analysis. Galaxy sizes are taken from the \citet{Meert15} catalog, which provides improved structural measurements, while stellar masses (\(\mathrm{M}_\ast\)) are obtained from the MPA/JHU value-added catalog\footnote{\url{http://www.mpa-garching.mpg.de/SDSS/DR7/}}.
Together, these datasets offer a reliable foundation for investigating the SMR across diverse environments.

Our study examines environments at the extremes of density, focusing on well defined high-density massive clusters and isolated low-density voids, to reassess the environmental impact on the SMR. High-density environments are identified using the GalWeight technique to assign galaxy memberships within clusters, while field galaxies located in voids represent very low-density regions. This membership-based approach minimizes environmental misclassifications, an inherent limitation of statistical density thresholds, and provides a robust framework for comparing galaxy populations across distinct environments.
To account for intrinsic differences in galaxy properties, we classify galaxies based on three independent criteria: star formation activity (quiescent, intermediate or star-forming), optical color (red, intermediate or blue), and morphology, as indicated by the bulge-to-total (B/T) light ratio (bulge-dominated, intermediate or disk-dominated). These classifications reflect key physical characteristics that may respond differently to environmental influences. For simplicity, we refer to quiescent, red or bulge-dominated galaxies as early-type, and to star-forming, blue or disk-dominated galaxies as late-type throughout this work.

In addition to presenting our own analysis, a key objective of this study is to synthesize and critically compare previous investigations of the SMR across environments. We compile a summary of earlier results, emphasizing differences in methodology, including the use of photometric versus spectroscopic data, approaches to measuring stellar mass and size, and definitions of environment. By contrasting these studies, we aim to identify the main sources of discrepancy and determine whether the reported environmental dependence of the SMR stems from physical effects or methodological biases. 
This paper is structured as follows. Section~\ref{sec:data} describes the data and methodology. Section~\ref{sec:results} presents our main results. In Section~\ref{sec:comparison}, we review previous studies of the SMR across environments, aiming to identify the sources of discrepancy in the literature. 
Finally, we summarize our results and conclusions in Section~\ref{sec:conc}. 
Throughout this paper, we adopt a flat $\Lambda$CDM cosmology consistent with the \citet{Planck15} results, assuming $\Omega_{\mathrm{M}} = 0.3089$, $\Omega_{\Lambda} = 0.6911$, and $h = 0.6774$. We use the term `log' to refer to base-10 logarithms (i.e., $\log_{10}$) throughout.
\section{Data and Methodology}\label{sec:data}
In this section, we describe the dataset used in our analysis to investigate the SMR of galaxies across different environments and galaxy populations. We also describe the methodology used to fit the SMR and derive its key parameters.

Our sample is drawn from the Sloan Digital Sky Survey Data Release 13 (SDSS--DR13; \citealp{Albareti17}). 
Key galaxy properties, including stellar mass (\(\mathrm{M_\ast}\)), star formation rate (SFR), rest-frame color ($g-r$), r-band absolute magnitude ($M_r$), half-light radius ($R_{\mathrm{tot}}$), half-light semi-major axis (\(R_{a}\)), and bulge-to-total (B/T) ratio, are compiled from a set of publicly available catalogs (Section~\ref{sec:properties}).
We classify galaxies into three populations based on: (i) their SFRs, (ii) optical colors, and (iii) morphologies (Section~\ref{sec:galclass}). Finally, we define the environments of interest by distinguishing between galaxies residing in clusters and those located in voids, enabling us to examine the impact of environment on galaxy structural properties (Section~\ref{sec:env}).

To construct our sample, we extract both photometric and spectroscopic data from SDSS-DR13 for galaxies that meet the following criteria. We include only objects classified as galaxies by the SDSS automated pipeline. We restrict the analysis to DR13 galaxies that have counterparts in the NYU Value-Added Galaxy Catalog (NYU-VAGC; SDSS-DR7; \citealp{Blanton05}) to ensure uniform ancillary photometric and derived quantities.
The sample is limited to galaxies with Petrosian $r$-band magnitudes in the range $13.9 \leq r \leq 17.6$, corrected for Galactic extinction following the method described by \citet{Blanton03a}. We further require that the spectroscopic redshift warning flag (\texttt{SpecObj.zWarning}) equals zero, indicating a reliable redshift measurement. In addition, we restrict our sample to galaxies with redshifts $z \leq 0.125$ as we discuss in Section \ref{sec:env}.

\begin{figure}
    \centering
    \includegraphics[width=1\linewidth]{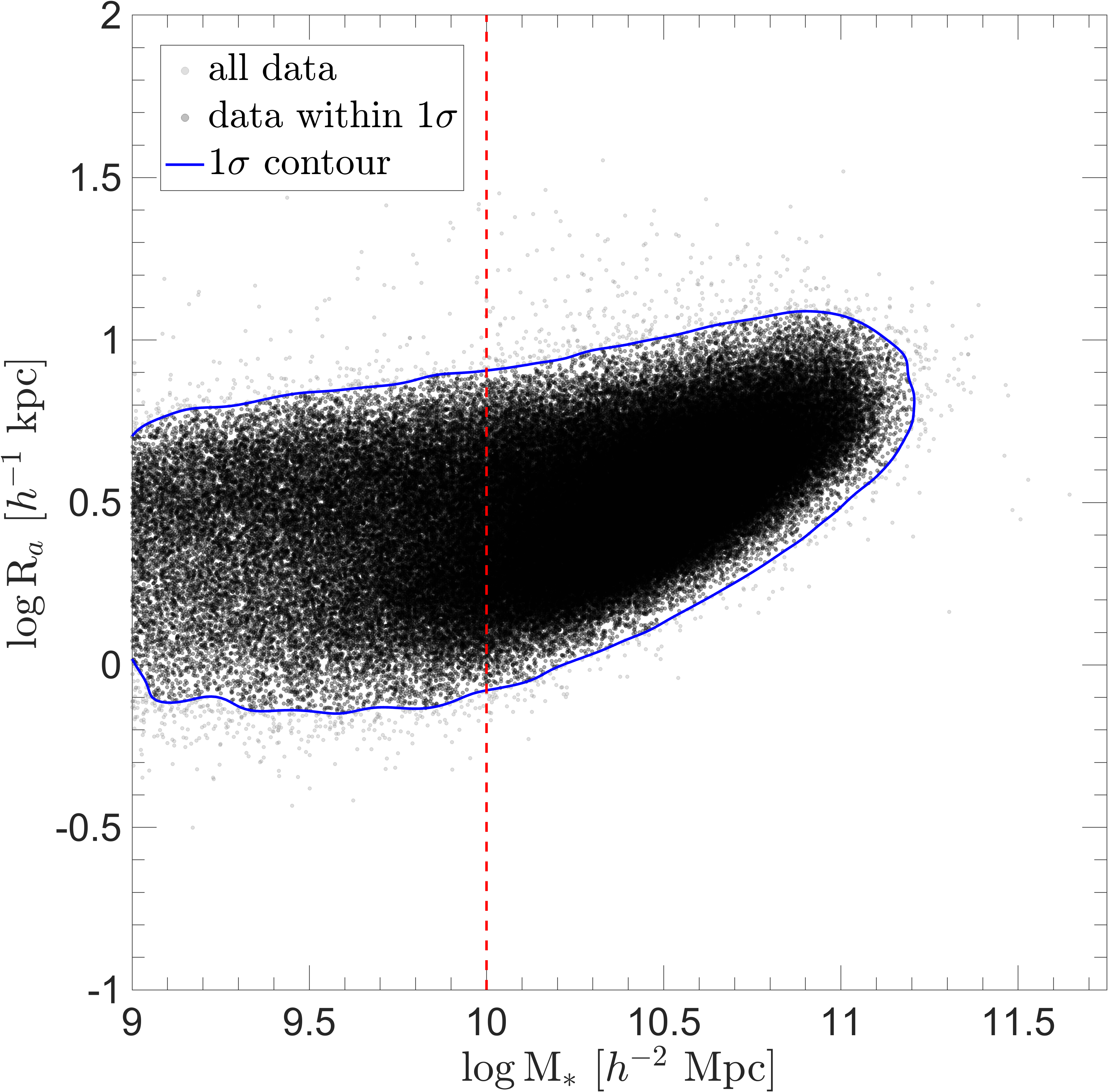} 
\caption{Distribution of SDSS galaxies in the size–stellar mass plane. Black points represent galaxies located within the $2\sigma$ contour (blue solid line). The red dashed line marks $\log{\mathrm{M}_\ast} = 10$~[\hmm] above which we perform the fitting (see Section \ref{sec:fitresults}).
}
    \label{fig:Fig01}
\end{figure}
\subsection{Galaxy Physical and Structural Properties} \label{sec:properties}

Both the stellar masses (\(\mathrm{M_\ast}\)) and star formation rates (SFRs) of galaxies are obtained from the MPA/JHU catalog. The stellar masses were calculated following the methodologies described in \citet{Kauffmann03} and \citet{Salim07}, using the \citet{Bruzual03} stellar population synthesis models and assuming a \citet{Kroupa01} initial mass function. It is worth noting that $\mathrm{M_\ast}$ estimates in the MPA/JHU catalog have been verified to be consistent with other independent measurements \citep[e.g.,][]{Taylor11,Leslie16}. 

To account for the SDSS flux-limited selection, we weight each galaxy by $1/V_{\max}$ \citep[e.g.,][]{Weigel16,Weaver23} in our binned measurements and in the SMR fitting, where $V_{\max}$ is the maximum comoving volume over which the galaxy would satisfy the survey magnitude limits. We present the SMR over the range $\log{M_\ast} \geq 9$~[\hmm], but we restrict our interpretation and conclusions to $\log{M_\ast} \geq 10$~[\hmm] as discussed later in this section. It is well known that stellar mass completeness is population dependent in flux-limited surveys, with blue (star-forming) galaxies remaining detectable to lower stellar masses than red (quiescent) galaxies at a given redshift because of their lower mass-to-light ratios \citep[e.g.,][]{Bosch08,Mosleh2018}. We account for this selection effect by applying $1/V_{\max}$ weighting, a standard non-parametric correction for the SDSS flux limit. Using the same SDSS dataset and the same $1/V_{\max}$ implementation, \citet{Paez24} showed that this approach robustly recovers the stellar mass function for the full population as well as for quiescent and star-forming subsamples out to $z\le 0.2$ (see their Section~4.3 and Figure~13).

SFRs are calculated using the nebular emission lines within the spectroscopic fiber aperture of 3 arcsec as described in \citet{Brinchmann04}. SFRs outside of the fiber are estimated using the galaxy photometry following \citet{Salim07}. SFRs are calculated using the empirical calibration of $\text{H}_\alpha$ emission lines \citep{Kennicutt98} and corrected from the dust extinction with the Balmer decrement $\text{H}_\alpha/\text{H}_\beta$ \citep{Charlot00}, assuming a Kroupa initial mass function \citep{Kroupa01}.


{\bf{
}}


Galaxy sizes are obtained from the catalog of \citet{Meert15}, which provides improved photometric measurements for 670,722 galaxies from SDSS. The catalog includes enhanced background subtraction techniques \citep{Vikram10,Bernardi13,Meert13,Bernardi14} and refined two-dimensional light profile fitting methods. In particular, galaxy sizes are derived by fitting multiple models to the $r$-band images, resulting in more accurate and consistent size estimates. 
To ensure robust structural measurements, we apply fit-quality cuts to the \citet{Meert15} models following the guidance in their Table~1. In particular, we use the \texttt{finalflag} bitmask to exclude unreliable fits and known failure modes that can arise in automated single- and two-component fitting (e.g., non-physical sizes or ellipticities and occasional component interchange in composite models). In our fiducial analysis we remove all objects flagged as either \emph{problematic} or \emph{bad} by requiring \texttt{finalflag} bits $<14$. In addition, to avoid size measurements dominated by seeing, we require the half-light radius of the total fit  to exceed the SDSS $r$-band PSF FWHM, i.e., $R_{\mathrm{tot}} > \mathrm{FWHM}_{\mathrm{PSF},r}$.

Throughout our analysis, we adopt the de~Vaucouleurs+Exponential (deVExp) half-light radius ($R_{\mathrm{tot}}$) as our fiducial size estimate. We obtain the same environmental trends when using the S\'ersic+Exponential (SerExp) size instead.
To adopt a consistent size definition, we compute the half-light semi-major axis radius as
\begin{equation}\label{eq:rad}
R_{a} = \frac{R_{\mathrm{tot}}}{\sqrt{ba_{\mathrm{tot}}}},
\end{equation}
where $ba_{\mathrm{tot}}$ is the axis ratio (minor-to-major axis, $b/a$) obtained from the total light profile fit \citep{Rodriguez21}. We have verified that our main conclusions are unchanged when using $R_{\mathrm{tot}}$ instead of $R_a$, and when restricting the sample to galaxies with $b/a>0.5$, indicating that our results are not sensitive to projection-related systematics.

Because galaxy size uncertainties are not provided in \citet{Meert15} for the adopted size measurements, we assign a simple fractional size uncertainty and propagate it to log space. In our fiducial analysis we assume a 5\% fractional uncertainty in size, $\Delta R_a = 0.05\,R_a$ (i.e., $R_a/20$), and include this term in  the likelihood together with the intrinsic-scatter component (Section~\ref{sec:fitting}). We verified that the inferred intrinsic scatter and our conclusions are stable under reasonable variations of this assumption (e.g., adopting larger fractional uncertainties), and therefore our results are not sensitive to the specific choice of size-error prescription.

\begin{figure*}
    \centering
    \includegraphics[width=1\linewidth]{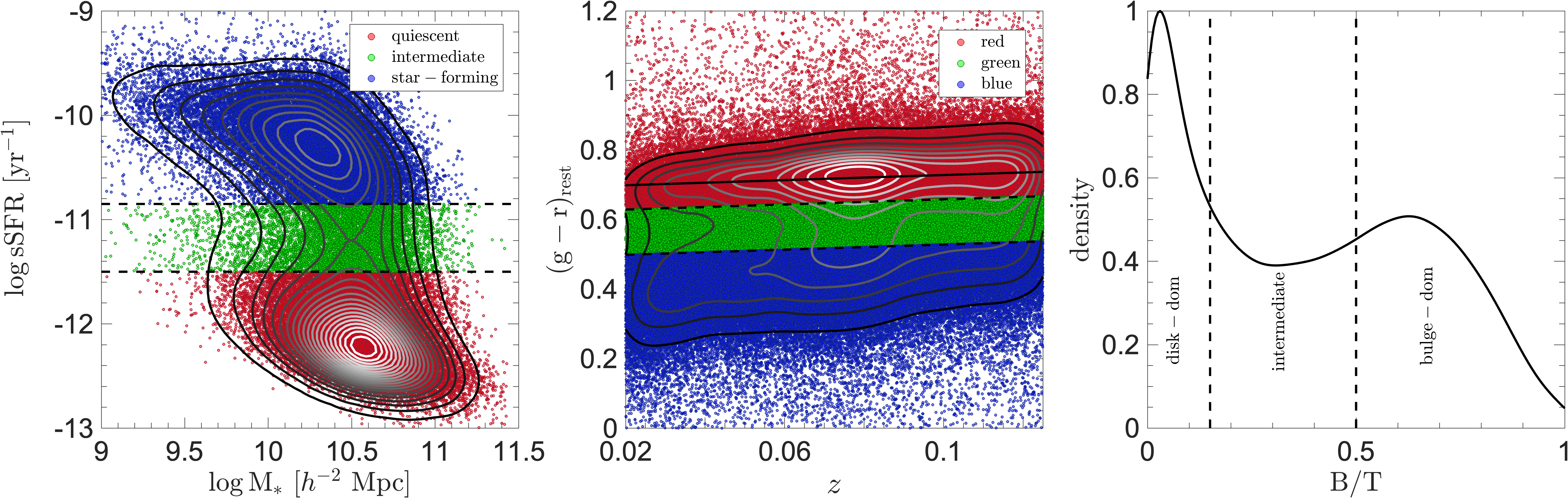} 
\caption{
Classification of galaxies. Left panel: Specific star formation rate (sSFR) as a function of stellar mass for cluster and void galaxies. Contour lines highlight the bimodal distribution, with peaks corresponding to star-forming (upper region) and quiescent (lower region) populations. Two horizontal dashed lines at $\mathrm{sSFR} = -11.50$ and $\mathrm{sSFR} = -10.85$ ($\sim 2\sigma$ from the two peaks) divide galaxies into quiescent, intermediate, and star-forming classes. 
Middle panel: Rest-frame $\mathrm{g{-}r}$ color as a function of redshift. The red-sequence ridgeline is identified in discrete redshift bins using a two-dimensional adaptive kernel density estimation method (see Section~\ref{sec:galclass}). The black solid line traces the red-sequence ridgeline across redshift, while the two dashed lines—offset by 0.07 mag ($\sim1\sigma$) and 0.2 mag ($\sim2\sigma$) below the ridgeline—define the color-based classification boundaries. Galaxies above the $1\sigma$ line are classified as red, those between the $1\sigma$ and $3\sigma$ lines as green, and those below the $3\sigma$ line as blue.
Right panel: Distribution of bulge-to-total (B/T) light ratios for the galaxy sample. Two vertical dashed lines at $B/T = 0.15$ and $B/T = 0.5$ ($\sim 2\sigma$ from the two peaks) divide galaxies into disk-dominated ($B/T < 0.15$), intermediate ($0.15 \leq B/T \leq 0.5$), and bulge-dominated ($B/T > 0.5$) classes.}
    \label{fig:class}
\end{figure*}

We take the $g$- and $r$-band total magnitudes from the \citet{Meert15} photometric catalog and use them to compute galaxy colors. After applying Galactic extinction corrections and $K$- and evolution-corrections to $z=0.1$, the resulting $(g-r)$ colors are rest-frame and are used for our color-based classification (Section~\ref{sec:galclass}).
Galaxy morphology is quantified using the bulge-to-total (B/T) light ratio, also taken from \citet{Meert15}, who performed two-dimensional bulge–disk decompositions on SDSS $r$-band images. In their modeling, the bulge component is fit with a de Vaucouleurs profile and the disk with an exponential profile. This decomposition yields a reliable structural classification based on light distribution, enabling us to distinguish between bulge-dominated and disk-dominated galaxies.

We cross-match the data from the different catalogs described above to obtain a unified set of galaxy properties for each object. This results in a sample of $\sim150,000$ galaxies with reliable measurements of $\mathrm{M_\ast}$, SFR, $R_{a}$, B/T ratio, and $g$- and $r$-band magnitudes. 
Figure~\ref{fig:Fig01} shows the distribution of the full galaxy sample in the size--stellar mass plane (gray points). 
To ensure an unbiased analysis, we apply an additional selection based on data density. Specifically, we include only galaxies (black points) enclosed within the $1\sigma$ contour level (blue line). This selection yields a sample of $\sim147,800$ galaxies with stellar masses of $\log{\mathrm{M}_\ast} \geq 9$~[\hmm].
We further restrict our interpretation and conclusions to galaxies with stellar masses $\log{M_\ast}\ge 10$~[$h^{-2}M_\odot$]. Figure~\ref{fig:Fig01} also shows that the size--mass plane becomes increasingly sparse below this threshold, with underpopulated regions that can be dominated by small-number statistics. At lower stellar masses, the SMR (particularly for early-type galaxies) appears to follow a broken power-law behavior. While some studies attribute the deviation at lower stellar masses to data incomleteness and selection effects, others suggest it may indicate an intrinsic change in the underlying relation \citep{Kauffmann2004}. For these reasons, we adopt $\log{M_\ast}\ge 10$~[$h^{-2}M_\odot$] for our main results \citep[e.g.,][]{Trujillio2011,Mosleh2018}.
\subsection{Galaxy Classification} \label{sec:galclass}
To investigate how the SMR varies with galaxy type, we divide our sample into three populations. This classification is based on three independent criteria: specific star formation rate (sSFR), optical color, and morphology (B/T ratio).

First, we begin by classifying galaxies according to their sSFR. As shown in Figure~\ref{fig:class} (left panel), the sSFR distribution exhibits a clear bimodal pattern, with two peaks corresponding to star-forming (upper peak) and quiescent (lower peak) galaxies, and a third, intermediate population lying between them. These groups are defined using two thresholds: $\log{\mathrm{sSFR}} = -11.50~[\mathrm{yr}^{-1}]$ and $\log{\mathrm{sSFR}} = -10.85~[\mathrm{yr}^{-1}]$. Galaxies with $\log{\mathrm{sSFR}} \leq -11.50$ are classified as quiescent, those with $\log{\mathrm{sSFR}} \geq -10.85$ as star-forming, and those with $-11.50 < \log{\mathrm{sSFR}} < -10.85$ as intermediate.

Second, we classify galaxies as red, green, or blue based on their optical colors using the red-sequence region in the color–magnitude diagram (CMD), following the method described in \citet[][Appendix A]{Abdullah23}. The red sequence is primarily populated by red, early-type galaxies that occupy a narrow region in the CMD, commonly referred to as the red sequence ridgeline \citep[e.g.,][]{Gladders00}. The position and slope of this ridgeline evolve smoothly with redshift \citep{Koester07}. 
To perform the classification, we divide the galaxy sample into 10 redshift bins of width $\Delta z \approx 0.01$. Within each bin, we construct the rest-frame $(\mathrm{g{-}r})_0$ vs. $M_r$ CMD and identify the red sequence ridgeline as the peak of the color distribution using a two-dimensional adaptive kernel method. We track the redshift evolution of this ridgeline, shown as the black solid line in Figure~\ref{fig:class} (middle panel).
We then define two color thresholds relative to the ridgeline: one at 0.07 mag below it  and another at 0.20 mag below it. Galaxies lying above the first line are classified as red; those between the first and second lines are classified as green; and galaxies below the second line are classified as blue. 

Third, to classify galaxies based on morphology, we use B/T ratio. As shown in Figure \ref{fig:class} (right panel) we apply the one-dimensional adaptive kernel method to the B/T distribution and identify a bimodal structure, with two prominent peaks separated by a local minimum. Based on this distribution, we define two thresholds at $B/T = 0.15$ and $B/T = 0.50$ to separate the sample into three morphological classes: disk-dominated galaxies with $B/T \leq 0.15$, bulge-dominated galaxies with $B/T \geq 0.50$, and intermediate systems in between.
\subsection{Environmental Classification} \label{sec:env}
To study the environmental effects on SMR, we define two extreme environments: galaxy clusters (high-density regions) and voids (low-density regions). 

Cluster galaxies are identified using the \texttt{GalWCat19} catalog \citep{Abdullah20a}, which assigns galaxies to dense environments based on precise membership determination. Galaxy clusters in this catalog are identified using the well-known Finger-of-God (FoG) effect \citep{Jackson72,Kaiser87,Abdullah13}. Cluster membership is assigned by applying the GalWeight technique \citep{Abdullah18} to galaxies in phase space, considering a projected radius of up to $10~h^{-1}$~Mpc and a line-of-sight velocity range of $\pm3000$~km~s$^{-1}$. In \citet{Abdullah18}, using the Bolshoi simulation \citep{Klypin16}, it was shown that GalWeight achieves $\sim98\%$ accuracy in identifying cluster members. 
The \texttt{GalWCat19} catalog contains two tables: one describing the properties of the clusters, and the other listing their member galaxies. The cluster table includes 1,800 clusters with redshifts in the range $0.01 \leq z \leq 0.2$ and total masses between $(0.4 - 14) \times 10^{14}~h^{-1}~M_{\odot}$. 
The galaxy member table includes 34,471 galaxies located within the virial radius of each cluster, defined as the radius where the density is 200 times the critical density of the Universe. 
To address the incompleteness issue in $\mathtt{GalWCat19}$ we select all clusters (and galaxies) with the redshifts $z \leqslant 0.125$ and cluster masses $\mathrm{M}_{200} \geq 13.9~[h^{-1}~\mathrm{M}_{\odot}]$ (see, \citealp{Abdullah20b}).

Void galaxies, representing low-density environments, are identified using a cylindrical method. For each galaxy in the sample, we define a cylinder centered on that galaxy, with a radius of $2~h^{-1}$~Mpc and a velocity depth of $3000~\mathrm{km~s^{-1}}$. We then count the number of galaxies within this cylinder, including the central one. If the total number of galaxies inside the cylinder is $\leq 8$ , we classify the central galaxy as a void galaxy. Galaxies identified in this way are considered to reside in low-density environments and are used to represent the field population in our analysis.
To ensure the robustness of our results, we investigate the potential systematic effects of our void definition on the SMR. In ~\ref{appendix:voids}, we examine how the SMR of galaxies in voids changes when varying the number of galaxies used to define the void environment. Specifically, we repeat our analysis using alternative thresholds, identifying void galaxies as those located at the center of cylinders containing 1, 5, 8, or 15 galaxies (including the central one). This test allows us to assess the sensitivity of our results to the adopted void selection criteria. We find that the inferred SMR and our conclusions are unchanged across these void definitions. We also tested an alternative low-density selection based on a nearest-neighbor metric and obtained consistent results.
\subsection{Fitting the Data}\label{sec:fitting}

In this section, we describe the methodology used to fit the SMR and derive its key parameters. We assume a lognormal distribution to model the probability of a galaxy having size \(\mathrm{R_{a}}\) at a given stellar mass \(\mathrm{M_\ast}\). The probability distribution is expressed as \citep{saro2015,simet2017}:
\begin{equation} \label{eq:prob}
\begin{split}
P(\log{\mathrm{R_a}}|\mathrm{M_\ast}) = ~~~~~~~~~~~~~~~~~~~~~~~~~~~~~~~~~~~~~~~~\\ \frac{1}{\sqrt{2\pi\sigma^2_{\log{\mathrm{R_a}},\mathrm{M_\ast}}}} \exp{\left[-\frac{\left(\log{\mathrm{R_a}} - \left<\log{\mathrm{R_a}}|\mathrm{M_\ast}\right>\right)^2}{2\sigma^2_{\log{\mathrm{R_a}},\mathrm{M_\ast}}}\right]},
\end{split}
\end{equation}
where \(\left<\log{\mathrm{R_a}}|\mathrm{M_\ast}\right>\) is the mean size at a given stellar mass, modeled as:
\begin{equation} \label{eq:rich}
\left<\log{\mathrm{R_a}}|\mathrm{M_\ast}\right> = \alpha + \beta \log{\left(\frac{\mathrm{M_\ast}}{\mathrm{M_{piv}}}\right)},
\end{equation}
where, $\log{\mathrm{M_{piv}}}= 10.5$ [\hmm]~and \(\sigma^2_{\log{\mathrm{R_a}},\mathrm{M_\ast}}\) represents the total variance in galaxy size at a fixed stellar mass, accounting for measurement uncertainties in both size (\(\sigma_{\log{\mathrm{R_a}}}\)) and stellar mass (\(\sigma_{\log{\mathrm{M_\ast}}}\)).

The total variance, including the intrinsic scatter in size (\(\sigma_{int}\)), is given by:
\begin{equation} \label{eq:var}
\sigma^2_{\log{\mathrm{R_a}},\mathrm{M_\ast}} = \beta^2\mathrm{M_\ast}^2 + \sigma^2_{\log{\mathrm{R_a}}} + \sigma^2_{int},
\end{equation}
where \(\beta\) denotes the slope of the SMR, and \(\alpha\) refers to the normalization. 
We neglect the impact of evolutionary effects in the SMR estimation due to the narrow redshift range of our data ($0.02 \leq z \leq 0.125$). 
To estimate the model parameters (\(\alpha\), \(\beta\), and \(\sigma_{\mathrm{int}}\)), we employ the affine-invariant Markov Chain Monte Carlo (MCMC) sampler from \citet{Goodman10}, implemented in the MATLAB package GWMCMC\footnote{\url{https://github.com/grinsted/gwmcmc}}, which is based on the Python package $\mathtt{emcee}$ \citep{Foreman13}.
\section{Results} \label{sec:results}
In this section, we present the SMR for galaxies in different environments and for various galaxy types, with the aim of assessing the influence of environment on this relation. 

\begin{figure}\centering
\includegraphics[width=1\linewidth]{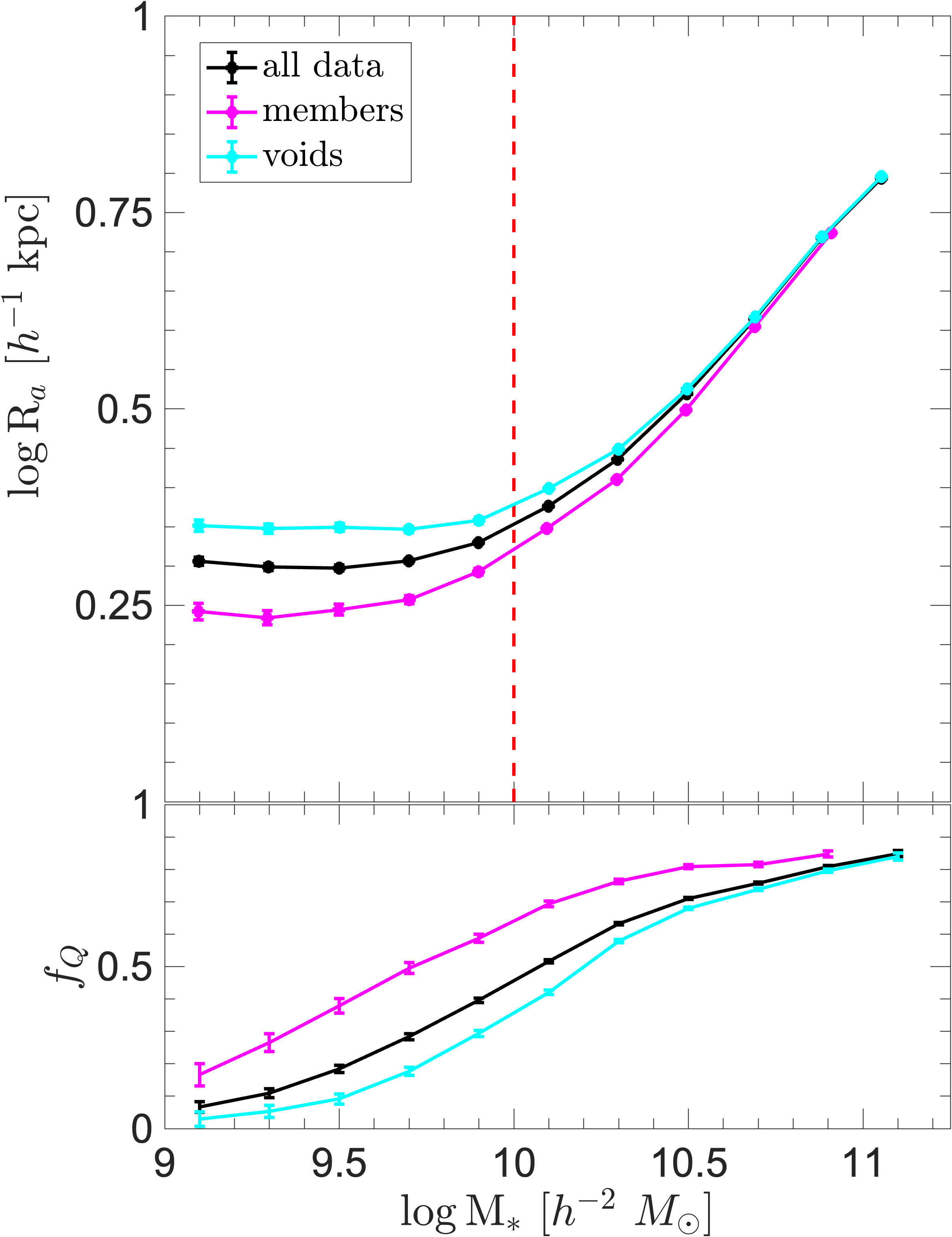} 
    \vspace{-0.5cm}
    \caption{Top: The size--stellar mass relation (SMR) for SDSS galaxies (black), $\mathtt{GalWCat19}$ cluster members (magenta), and void galaxies (cyan). Error bars show 1$\sigma$ standard errors on the $1/V_{\max}$-weighted mean of $\log{R_a}$ in each stellar mass bin, computed as $\sigma=\sigma_w/\sqrt{N_{\rm eff}}$, where $\sigma_w$ is the weighted scatter and $N_{\rm eff}=(\sum w)^2/\sum(w^2)$ is the effective sample size. Bottom: the quiescent-galaxy fraction, $f_Q\equiv N_Q/N_{\rm tot}$, as a function of stellar mass for the same samples and bins, with binomial uncertainties $\sigma_{f_Q}=\sqrt{f_Q(1-f_Q)/N_{\rm tot}}$.
    }
    \label{fig:All}
\end{figure}

\subsection{Size--Mass Relation for Cluster and Void Galaxies} \label{sec:resultsa}

Figure~\ref{fig:All} presents the SMR for the entire galaxy population (black solid line). The results show a clear and strong correlation, where galaxy size increases systematically with stellar mass. This trend is consistent with the well-established scaling relations reported in previous studies \citep[e.g.,][]{Shen03,Lange15}, indicating that larger galaxies tend to have higher stellar masses. Such a relation reflects the hierarchical nature of galaxy growth and highlights the role of both internal processes, such as star formation, and external mechanisms, including mergers and accretion. 

Figure~\ref{fig:All} also compares the SMR across two extreme environments: cluster member galaxies identified in the $\mathtt{GalWCat19}$ catalog (solid magenta line) and void galaxies (solid cyan line). In the top panel, void galaxies exhibit slightly larger sizes than cluster members at the low-mass end, although the overall trends are broadly similar. The bottom panel of Figure~\ref{fig:All} helps interpret this difference by showing the quiescent-galaxy fraction, $f_Q \equiv N_Q/N_{\rm tot}$, as a function of stellar mass in each environment. We find that $f_Q^{\rm mem}(M_\star) > f_Q^{\rm void}(M_\star)$ at fixed stellar mass, especially toward the low-mass end, implying that the two environments contain different mixtures of galaxy types. Since quiescent and star-forming galaxies follow different SMRs, combining all galaxy types can produce an apparent offset between environments. Thus, the apparent offset in the all-galaxy SMR is driven primarily by differences in the quiescent fraction (population mix), rather than a change in the size--mass relation at fixed galaxy type. This motivates our subsequent analysis, where we compare the SMR at fixed population (quiescent, intermediate, and star-forming) to isolate any environmental dependence.

\begin{figure*}\centering
\includegraphics[width=1\linewidth]{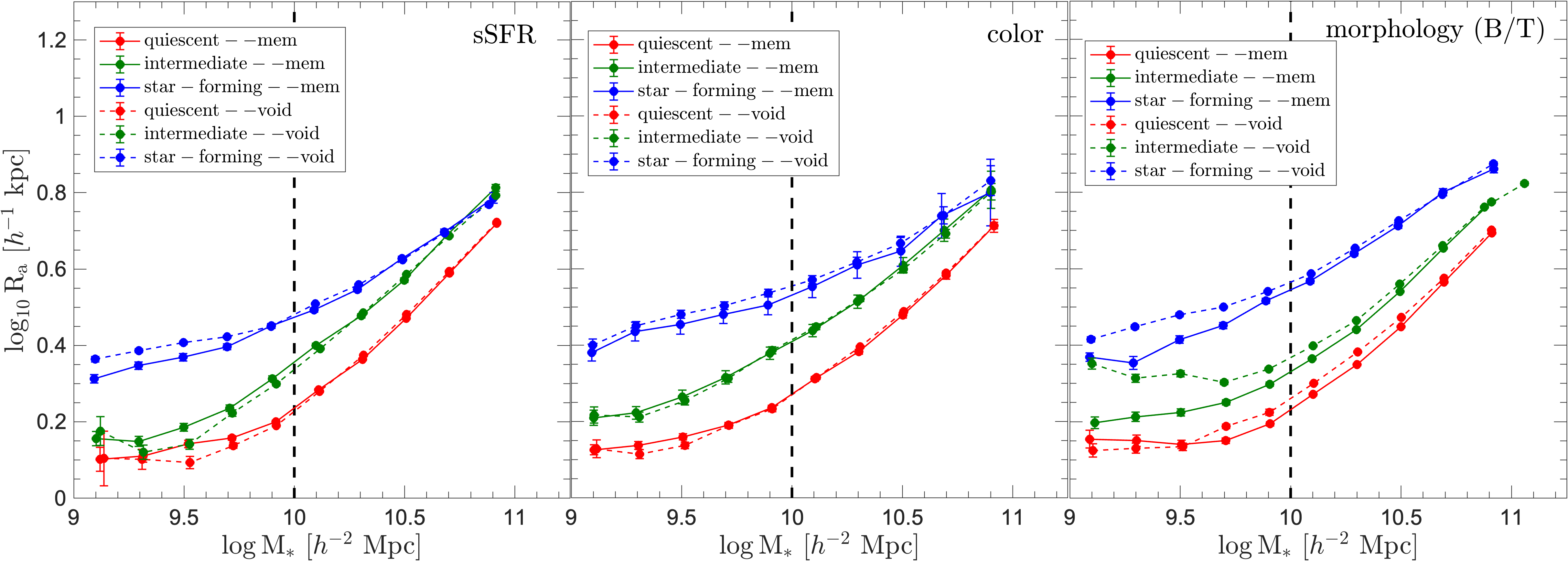} 
    \vspace{-0.2cm}
    \caption{The size-stellar mass relation for galaxies divided by galaxy type and environment. Solid lines represent cluster galaxies, while dashed lines correspond to void galaxies. Colors indicate galaxy types: early (red), intermediate (green), and late (blue) as shown in the legend. Galaxies are classified based on: sSFR (left), color (middle), and B/T (right). The SMR for each type is consistent across cluster and void environments, indicating no environmental effects on the SMR within each type.
    }
    \label{fig:membersvoids}
\end{figure*}

In this analysis, we classify both cluster and void galaxy populations into three distinct types based on sSFR activity, optical color, and B/T ratio as described in Section~\ref{sec:galclass}. Figure~\ref{fig:membersvoids} displays the SMR for these classifications in three separate panels: left (sSFR-based), middle (color-based), and right (B/T-based). In each panel, we compare cluster galaxies—early (solid red), late (solid blue), and intermediate (solid green)—with their void counterparts—early (dashed red), late (dashed blue), and intermediate (dashed green). The figure reveals four main findings, which we discuss below. Importantly, our conclusions remain consistent across all classification types, indicating that the choice of method used to separate galaxy types does not impact the observed trends.

First, we find no significant evidence for environmental effects on the SMR. The SMR trends for corresponding galaxy types in clusters and voids are remarkably consistent across all classification types. This indicates that, when galaxies are grouped by intrinsic properties such as star formation activity, color, or morphology, environmental density has no significant impact on their structural scaling relations. This consistency aligns with previous studies \citep[e.g.,][]{Weinmann2009,Huertas-Company2013a}, which have also reported minimal environmental influence on galaxy scaling relations when intrinsic properties are accounted for. These results support the view that galaxy size is primarily governed by internal processes such as star formation, feedback, and initial conditions rather than by large scale environmental factors.

Second, early-type galaxies exhibit steeper SMR slopes compared to their late-type counterparts. This suggests that the size growth of early-type galaxies is more strongly correlated with stellar mass. This result is consistent with previous studies \citep[e.g.,][]{Shen03,vanderwel2014}, which also reported steeper SMR slopes for early-type systems. The steeper slope reflects the compact morphologies and distinct evolutionary pathways of early-type galaxies, shaped by mechanisms such as mergers and internal dynamical instabilities processes like bar formation, disk instabilities, and violent relaxation that act within galaxies to redistribute mass and build central structures \citep{Naab09,Hopkins09,Frigo17}.

\begin{figure*}
    \centering    \includegraphics[width=1\linewidth]{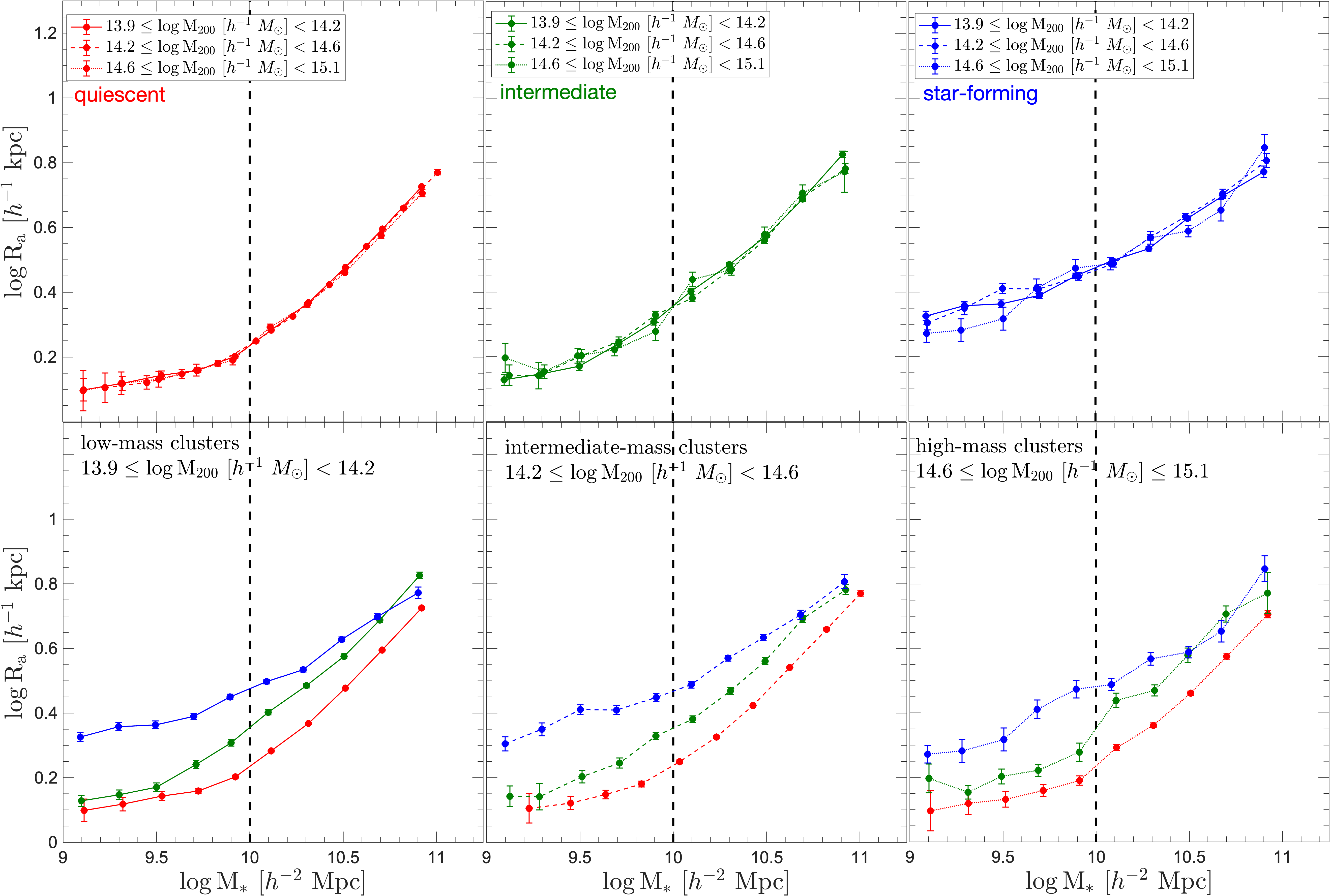} 
    \vspace{-0.2cm}
    \caption{Upper panels: Size-stellar mass relation (SMR) for galaxies in $\mathtt{GalWCat19}$ clusters, divided into three mass subsample: $13.9 \leq\log\mathrm{M_{200}} < 14.2~[h^{-1}\mathrm{M_{\odot}}]$ (solid lines), $14.2 \leq \log\mathrm{M_{200}} < 14.6~[h^{-1}~\mathrm{M_{\odot}}]$ (dashed lines), and $14.6 \leq \log\mathrm{M_{200}} \leq 15.1~[h^{-1}~\mathrm{M_{\odot}}]$ (dotted lines). Panels show the SMR for quiescent (left), intermediate (middle), and star-forming (right) galaxies. Lower panels: SMR for the three galaxy types combined in each cluster mass bin. The plot indicates minimal differences in the SMR across the three cluster mass bins for each galaxy type.}
    \label{fig:mass}
\end{figure*}

Third, at lower stellar masses, late-type galaxies exhibit significantly larger sizes than early-type galaxies, with intermediate galaxies occupying a transitional position between the two. We rule out sky subtraction as a source of this trend, as \citet{Meert13} addressed this issue through dedicated corrections that improve background estimation in the data.
The larger sizes of late-type galaxies are attributed to their extended stellar disks and lower surface mass densities, maintained by ongoing star formation that preserves their disk structures \citep{Kauffmann03,Kauffmann2003b,Saintonge2011,vanderwel2014}. Intermediate galaxies, representing a transitional phase, retain partial disk components while undergoing star formation quenching, resulting in intermediate sizes \citep{Pandya17,Brennan2017}. 
In contrast, early-type galaxies exhibit more compact morphologies, shaped by gas exhaustion and structural transformations driven by mergers or internal dynamical processes \citep{Naab09,Hopkins09}.
These trends are consistent with previous studies, which have reported that late-type galaxies are systematically larger than early types at fixed stellar mass, and that early-type galaxies tend to quench earlier and host older stellar populations—reflecting differences in their evolutionary timescales and assembly histories \citep{vanderwel2014}.

Fourth, as stellar mass increases, the size differences among the three galaxy types progressively diminish, with all types converging to similar sizes at the high-mass end ($\log M_{\ast} \geq 11$\;[\hmm]). This convergence suggests that massive galaxies are predominantly shaped by intrinsic processes, such as mergers and structural compaction, rather than ongoing star formation. The growth in size for massive galaxies is consistent with findings from \citet{fan2008,Ryan2012,Rosito2021}, who attributed this phenomenon to AGN feedback and associated mass-loss mechanisms. Furthermore, this growth aligns with the hypothesis that massive galaxies experience distinct star formation histories compared to less massive systems \citep{trujillio2020}.


\begin{figure*}
    \centering
 \includegraphics[width=0.9\linewidth]{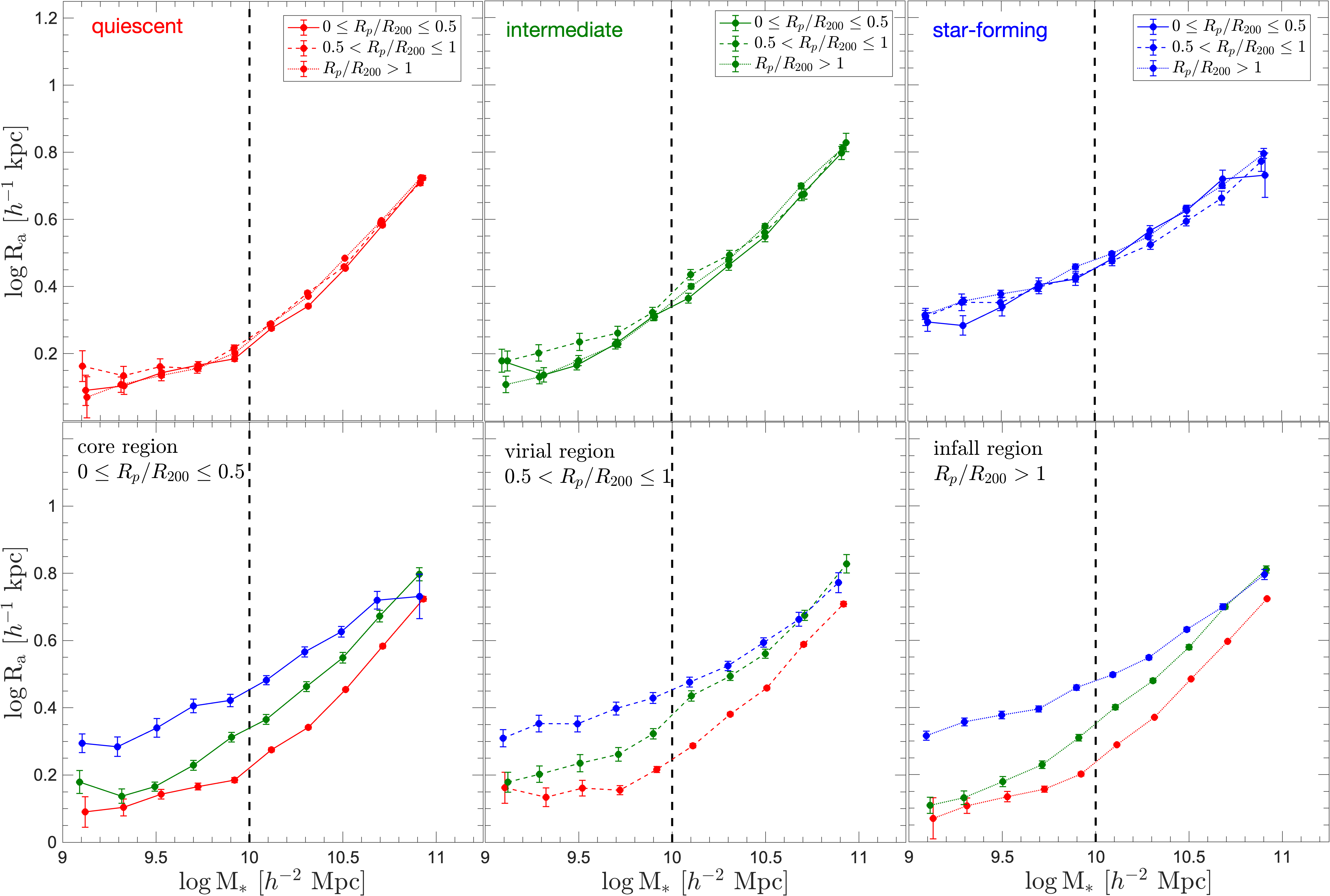} 
    \vspace{-0.25cm}
    \caption{Upper panels: Size-stellar mass relation (SMR) for galaxies in $\mathtt{GalWCat19}$ clusters, divided into three regions within clusters: \(0\leq \mathrm{R_{p}}/R_{200} < 0.5 \, \) (solid lines), \( 0.5 \,  \leq \mathrm{R_{p}}/R_{200} < 1 \) (dashed line), and \( \mathrm{R_{p}}/R_{200}  \geq 1\) (dotted line).
     Panels show the SMR for quiescent (left), intermediate (middle), and star-forming (right) galaxies. Lower panels: SMR for the three galaxy types combined in each cluster region. The plot indicates minimal differences in the SMR across the three cluster regions for each galaxy type.}
    \label{fig:radius}
\end{figure*}

\subsection{Dependence of the Size–Mass Relation on Cluster Mass} \label{sec:mass}
In this section, we examine how the SMR depends on cluster mass. To this end, we divide galaxies in the $\mathtt{GalWCat19}$ catalog into three subsamples based on the cluster virial mass ($\mathrm{M_{200}}$): $13.9 \leq \log\mathrm{M_{200}}~[h^{-1}~\mathrm{M_{\odot}}] < 14.2$, $14.2 \leq \log\mathrm{M_{200}}~[h^{-1}~\mathrm{M_{\odot}}] < 14.6$, and $14.6 \leq \log\mathrm{M_{200}}~[h^{-1}~\mathrm{M_{\odot}}] \leq 15.1$.

The upper three panels of Figure~\ref{fig:mass} show the SMR for quiescent (left), intermediate (middle), and star-forming (right) galaxies, classified according to their sSFR. Within each panel, we compare SMR trends for galaxies in low-mass (solid lines), intermediate-mass (dashed lines), and high-mass (dotted lines) clusters. The trends across all three mass bins are consistent for each galaxy type, indicating that cluster mass has little to no influence on galaxy sizes at fixed stellar mass.

The lower three panels of Figure~\ref{fig:mass} present the SMR for the three galaxy populations within each cluster mass bin, enabling a direct comparison of their sizes at fixed stellar mass. 
Regardless of cluster mass, star-forming galaxies consistently exhibit larger sizes than intermediate and quiescent galaxies at fixed stellar mass. These results suggest that the SMR is largely independent of the host cluster mass, reinforcing the idea that galaxy sizes are primarily governed by internal processes. The minimal variation across different environments implies that large-scale gravitational and dynamical properties of clusters play a secondary role, while internal mechanisms (such as star formation, feedback, and structural evolution) dominate in shaping the SMR.

We verify the robustness of our results by repeating the analysis using galaxy classifications based on color and B/T ratio, as shown in ~\ref{appendix:color_bt}. In both cases, we recover the same trends, reinforcing the conclusion that the observed SMR behavior is not sensitive to the specific classification method used. This suggests that the SMR is primarily governed by intrinsic galaxy properties, with minimal influence from environmental factors. This consistency across classifications and environments confirms the lack of significant environmental influence on the SMR.

\subsection{Dependence of the Size–Mass Relation on Cluster-Centric Distance} \label{sec:radius}

In this section, we explore how the SMR depends on galaxy location within clusters. To characterize the internal cluster environment, we divide galaxies in the $\mathtt{GalWCat19}$ catalog into three subsamples based on their projected cluster-centric radius \( \mathrm{R_p} \), normalized by the virial radius \( R_{200} \). Galaxies with \( 0 \leq \mathrm{R_p}/R_{200}< 0.5\) reside in the core region, those with \( 0.5 \, \leq \mathrm{R_p} /R_{200}< 1 \) occupy the virial region, and those with \(\mathrm{R_p}/R_{200} \geq 1 \) lie within the infall region.
The upper three panels of Figure~\ref{fig:radius} show the SMR for each galaxy type, quiescent (left), intermediate (middle), and star-forming (right), within the three radial bins. Solid, dashed, and dotted lines represent galaxies in the core, virial, and infall regions, respectively. The trends across all radial bins are remarkably consistent for each galaxy type, indicating no significant variation in the SMR as a function of cluster-centric distance. This supports our findings that the correlation between galaxy size and stellar mass is not influenced by local position within the cluster environment.

The lower three panels of Figure~\ref{fig:radius} display the SMR for all three galaxy types within each radial bin. The left panel compares quiescent, intermediate, and star-forming galaxies in the core region, the middle panel is in the virial region, and the right panel is in the infall region. Similar to our findings in Section~\ref{sec:mass}, we observe that at lower stellar masses, star-forming galaxies exhibit the largest sizes, followed by intermediate  galaxies, while quiescent galaxies are the most compact. This size hierarchy diminishes at higher stellar masses, where all three galaxy types converge to similar sizes.
To ensure our conclusions are not sensitive to classification method, we repeat the analysis using optical color and B/T ratio as alternative criteria, as shown in ~\ref{appendix:color_bt}. The conclusions remain unchanged. This consistency across radial regions and classification methods reinforces the finding that the SMR within clusters is primarily governed by intrinsic galaxy properties, with minimal influence from location within the cluster or external environmental effects.
\begin{table}
\centering
\caption{
Best-fit parameters of the size--stellar mass relation, $\left<\log{\mathrm{R_a}}|\mathrm{M_\ast}\right> = \alpha + \beta \log{\left(\mathrm{M_\ast}/\mathrm{M_{piv}}\right)}$, for galaxies in clusters and voids, where $\log{\mathrm{M_{piv}}} = 10.5$~[\hmm]. Results are shown for three galaxy classification types, early (E), intermediate (I), and late (L) based on specific star formation rate (sSFR), optical color, and bulge-to-total (B/T) ratio. Uncertainties represent the $1\sigma$ errors from the MCMC posterior distributions.
}
\label{tab:fitresults}
\begin{tabular}{llccc}
\hline
Class & Env & $\alpha$ & $\beta$ & $\sigma_{\mathrm{int}}$\\
\hline
\hline
sSFR  & E--mem   & $0.48 \pm 0.02$ & $0.53 \pm 0.03$ & $0.08 \pm 0.01$ \\
sSFR  & I--mem   & $0.58 \pm 0.02$ & $0.47 \pm 0.03$ & $0.10 \pm 0.01$ \\
sSFR  & L--mem   & $0.63 \pm 0.02$ & $0.34 \pm 0.03$ & $0.13 \pm 0.01$ \\
sSFR  & E--void  & $0.48 \pm 0.02$ & $0.56 \pm 0.03$ & $0.07 \pm 0.01$ \\
sSFR  & I--void  & $0.58 \pm 0.02$ & $0.51 \pm 0.03$ & $0.09 \pm 0.01$ \\
sSFR  & L--void  & $0.63 \pm 0.02$ & $0.31 \pm 0.03$ & $0.14 \pm 0.01$ \\
\hline\hline
color & E--mem   & $0.49 \pm 0.02$ & $0.48 \pm 0.03$ & $0.10 \pm 0.01$ \\
color & I--mem   & $0.62 \pm 0.02$ & $0.43 \pm 0.03$ & $0.11 \pm 0.01$ \\
color & L--mem   & $0.65 \pm 0.02$ & $0.27 \pm 0.03$ & $0.16 \pm 0.01$ \\
color & E--void  & $0.49 \pm 0.02$ & $0.49 \pm 0.03$ & $0.09 \pm 0.01$ \\
color & I--void  & $0.61 \pm 0.02$ & $0.41 \pm 0.03$ & $0.12 \pm 0.01$ \\
color & L--void  & $0.67 \pm 0.02$ & $0.26 \pm 0.03$ & $0.14 \pm 0.01$ \\
\hline\hline
B/T   & E--mem   & $0.46 \pm 0.02$ & $0.51 \pm 0.03$ & $0.08 \pm 0.01$ \\
B/T   & I--mem   & $0.55 \pm 0.02$ & $0.49 \pm 0.03$ & $0.10 \pm 0.01$ \\
B/T   & L--mem   & $0.72 \pm 0.02$ & $0.37 \pm 0.03$ & $0.09 \pm 0.01$ \\
B/T   & E--void  & $0.48 \pm 0.02$ & $0.49 \pm 0.03$ & $0.08 \pm 0.01$ \\
B/T   & I--void  & $0.57 \pm 0.02$ & $0.46 \pm 0.03$ & $0.12 \pm 0.01$ \\
B/T   & L--void  & $0.73 \pm 0.02$ & $0.34 \pm 0.03$ & $0.10 \pm 0.01$ \\
\hline
\end{tabular}
\end{table}

\begin{table*}
\centering
\caption{Same as Table~\ref{tab:fitresults}, but showing the best-fit SMR parameters for galaxies across two environmental classifications. The left set of columns presents results based on cluster mass (low, intermediate, and high), while the right set presents results based on cluster-centric distance (core, virial, and infall regions).}
\label{tab:massradius}
\begin{tabular}{l||lccc||lccc}
\hline
Class & Env (Mass) & $\alpha$ & $\beta$ & $\sigma_{\mathrm{int}}$ & Env (Distance) & $\alpha$ & $\beta$ & $\sigma_{\mathrm{int}}$ \\
\hline
sSFR & E--low   & $0.48 \pm 0.02$ & $0.54 \pm 0.03$ & $0.08 \pm 0.01$ & E--core   & $0.46 \pm 0.02$ & $0.53 \pm 0.03$ & $0.09 \pm 0.01$ \\
sSFR & I--low   & $0.59 \pm 0.02$ & $0.48 \pm 0.03$ & $0.10 \pm 0.01$ & I--core   & $0.56 \pm 0.02$ & $0.49 \pm 0.04$ & $0.11 \pm 0.01$ \\
sSFR & L--low   & $0.62 \pm 0.02$ & $0.33 \pm 0.03$ & $0.13 \pm 0.01$ & L--core   & $0.63 \pm 0.02$ & $0.34 \pm 0.04$ & $0.11 \pm 0.01$ \\
sSFR & E--inter & $0.47 \pm 0.02$ & $0.53 \pm 0.03$ & $0.08 \pm 0.01$ & E--virial & $0.48 \pm 0.02$ & $0.50 \pm 0.03$ & $0.08 \pm 0.01$ \\
sSFR & I--inter & $0.57 \pm 0.02$ & $0.48 \pm 0.03$ & $0.10 \pm 0.01$ & I--virial & $0.58 \pm 0.02$ & $0.40 \pm 0.04$ & $0.11 \pm 0.01$ \\
sSFR & L--inter & $0.64 \pm 0.02$ & $0.38 \pm 0.03$ & $0.13 \pm 0.01$ & L--virial & $0.60 \pm 0.02$ & $0.33 \pm 0.04$ & $0.12 \pm 0.01$ \\
sSFR & E--high  & $0.47 \pm 0.02$ & $0.49 \pm 0.03$ & $0.09 \pm 0.01$ & E--infall & $0.48 \pm 0.02$ & $0.54 \pm 0.03$ & $0.08 \pm 0.01$ \\
sSFR & I--high  & $0.59 \pm 0.03$ & $0.44 \pm 0.04$ & $0.13 \pm 0.01$ & I--infall & $0.59 \pm 0.02$ & $0.49 \pm 0.03$ & $0.10 \pm 0.01$ \\
sSFR & L--high  & $0.62 \pm 0.03$ & $0.32 \pm 0.04$ & $0.12 \pm 0.01$ & L--infall & $0.63 \pm 0.02$ & $0.35 \pm 0.03$ & $0.13 \pm 0.01$ \\
\hline\hline
color & E--low   & $0.49 \pm 0.02$ & $0.48 \pm 0.03$ & $0.10 \pm 0.01$ & E--core   & $0.47 \pm 0.02$ & $0.50 \pm 0.03$ & $0.10 \pm 0.01$ \\
color & I--low   & $0.62 \pm 0.02$ & $0.42 \pm 0.03$ & $0.11 \pm 0.01$ & I--core   & $0.61 \pm 0.02$ & $0.44 \pm 0.04$ & $0.12 \pm 0.01$ \\
color & L--low   & $0.65 \pm 0.02$ & $0.25 \pm 0.03$ & $0.16 \pm 0.01$ & L--core   & $0.57 \pm 0.03$ & $0.32 \pm 0.05$ & $0.18 \pm 0.01$ \\
color & E--inter & $0.48 \pm 0.02$ & $0.48 \pm 0.03$ & $0.10 \pm 0.01$ & E--virial & $0.49 \pm 0.02$ & $0.45 \pm 0.03$ & $0.10 \pm 0.01$ \\
color & I--inter & $0.61 \pm 0.02$ & $0.45 \pm 0.03$ & $0.11 \pm 0.01$ & I--virial & $0.60 \pm 0.02$ & $0.42 \pm 0.04$ & $0.11 \pm 0.01$ \\
color & L--inter & $0.67 \pm 0.02$ & $0.29 \pm 0.04$ & $0.16 \pm 0.01$ & L--virial & $0.63 \pm 0.03$ & $0.24 \pm 0.05$ & $0.14 \pm 0.01$ \\
color & E--high  & $0.48 \pm 0.02$ & $0.44 \pm 0.03$ & $0.11 \pm 0.01$ & E--infall & $0.50 \pm 0.02$ & $0.47 \pm 0.03$ & $0.10 \pm 0.01$ \\
color & I--high  & $0.61 \pm 0.03$ & $0.44 \pm 0.04$ & $0.12 \pm 0.01$ & I--infall & $0.62 \pm 0.02$ & $0.43 \pm 0.03$ & $0.11 \pm 0.01$ \\
color & L--high  & $0.66 \pm 0.03$ & $0.36 \pm 0.06$ & $0.18 \pm 0.02$ & L--infall & $0.68 \pm 0.02$ & $0.27 \pm 0.03$ & $0.16 \pm 0.01$ \\
\hline\hline
B/T   & E--low   & $0.46 \pm 0.02$ & $0.52 \pm 0.03$ & $0.08 \pm 0.01$ & E--core   & $0.44 \pm 0.02$ & $0.54 \pm 0.03$ & $0.08 \pm 0.01$ \\
B/T   & I--low   & $0.55 \pm 0.02$ & $0.49 \pm 0.03$ & $0.10 \pm 0.01$ & I--core   & $0.53 \pm 0.02$ & $0.49 \pm 0.03$ & $0.10 \pm 0.01$ \\
B/T   & L--low   & $0.72 \pm 0.02$ & $0.38 \pm 0.03$ & $0.09 \pm 0.01$ & L--core   & $0.69 \pm 0.02$ & $0.39 \pm 0.04$ & $0.06 \pm 0.01$ \\
B/T   & E--inter & $0.46 \pm 0.02$ & $0.51 \pm 0.03$ & $0.08 \pm 0.01$ & E--virial & $0.46 \pm 0.02$ & $0.49 \pm 0.03$ & $0.08 \pm 0.01$ \\
B/T   & I--inter & $0.55 \pm 0.02$ & $0.50 \pm 0.03$ & $0.10 \pm 0.01$ & I--virial & $0.54 \pm 0.02$ & $0.47 \pm 0.03$ & $0.09 \pm 0.01$ \\
B/T   & L--inter & $0.72 \pm 0.02$ & $0.35 \pm 0.04$ & $0.09 \pm 0.01$ & L--virial & $0.71 \pm 0.02$ & $0.39 \pm 0.04$ & $0.07 \pm 0.01$ \\
B/T   & E--high  & $0.45 \pm 0.02$ & $0.49 \pm 0.03$ & $0.09 \pm 0.01$ & E--infall & $0.47 \pm 0.02$ & $0.51 \pm 0.03$ & $0.08 \pm 0.01$ \\
B/T   & I--high  & $0.54 \pm 0.02$ & $0.45 \pm 0.04$ & $0.11 \pm 0.01$ & I--infall & $0.56 \pm 0.02$ & $0.49 \pm 0.03$ & $0.11 \pm 0.01$ \\
B/T   & L--high  & $0.72 \pm 0.03$ & $0.38 \pm 0.05$ & $0.09 \pm 0.01$ & L--infall & $0.72 \pm 0.02$ & $0.36 \pm 0.03$ & $0.09 \pm 0.01$ \\
\hline
\end{tabular}
\end{table*}

\subsection{Fitting Results of the Size–Mass Relation} \label{sec:fitresults}

In this section, we present the results of fitting the size–mass relation (SMR) for different galaxy populations, using the methodology outlined in Section~\ref{sec:fitting}. The SMR parameters—normalization ($\alpha$), slope ($\beta$), and intrinsic scatter ($\sigma_{\mathrm{int}}$)—are estimated via Markov Chain Monte Carlo (MCMC) sampling. We restrict our analysis to galaxies with stellar masses $\log{\mathrm{M_\ast}} \geq 10$~[$h^{-2}M_\odot$] as discussed in Section~\ref{sec:properties}.

Table~\ref{tab:fitresults} summarizes the best-fit parameters of the SMR for cluster and void galaxies, classified by sSFR, color, and B/T morphology. A key result from this table is the comparison between cluster members and void galaxies: for each galaxy type—early, intermediate, and late—the slope and normalization of the SMR are remarkably consistent across these two extreme environments. This consistency suggests that the large-scale environment, whether overdense or underdense, has minimal impact on the structural scaling relation of galaxies when stellar mass and internal properties are fixed. These findings support a scenario in which internal processes, such as star formation, feedback, and mergers, play the dominant role in driving the size evolution of galaxies. Moreover, the results confirm the expected trend: late-type galaxies (star-forming, blue, or disk-dominated) exhibit a shallower slope and higher normalization, reflecting their larger typical sizes at fixed stellar mass. In contrast, early-type galaxies (quiescent, red, or bulge-dominated) display steeper slopes and more compact sizes. Intermediate types fall between these two extremes. The intrinsic scatter is relatively small across all categories, indicating that the SMR is a well-defined, tight, and robust relation within each galaxy type. These fits align with the trends observed in the SMR plots presented earlier, reinforcing the reliability of the relation across different classification types and environments.

The left half of Table~\ref{tab:massradius} summarizes the best fit parameters of the size stellar mass relation for galaxies in clusters of different masses. Across all three classification types sSFR, color, and B/T we find that for each galaxy type (early, intermediate, and late), the slope ($\beta$) and normalization ($\alpha$) of the relation remain nearly constant across the low-, intermediate-, and high-mass cluster bins. This consistency indicates that the total mass of the host cluster does not have a significant impact on the structural scaling relations of galaxies once stellar mass and type are fixed.
The right half of Table~\ref{tab:massradius} shows the corresponding fits as a function of projected cluster-centric distance, divided into core, virial, and infall regions. As with the mass-based division, we find no significant variation in the SMR parameters across the three radial zones. For each galaxy type, the slope and normalization of the relation remain stable regardless of whether galaxies reside near the cluster center or in the outskirts. This suggests that local environment, as traced by distance from the cluster center, does not play a dominant role in shaping the size–mass relation.
Collectively, these findings provide evidence that the size of galaxies in the local universe is governed primarily by internal processes, with external factors such as cluster mass and galaxy position playing a secondary or negligible role.

\begin{table*}
\centering
\caption{Summary of studies investigating the environmental dependence of the galaxy size--mass relation. 
The table lists the redshift range, data quality, environmental definition, physical scale probed, stellar-mass range, and the reported trend for each study.
}
\label{tab:env}
\begin{tabular}{lcccccc}
\hline
Study & $z$ range & quality & Environment & Scale & $\log{M_\star}$~[\hmm] & Trend \\
\hline
\hline
\multicolumn{7}{c}{Environmental Independence} \\
\hline
This work                   & 0.01--0.125   & spec      & Cls/F        & LS & $\geq$10.00 & none \\
\cite{Afanasiev2023}        & 1.40--2.80    & spec      & Cls/F        & LS    & 9.30--10.20 & none \\
\cite{Zhang2019}            & 0.01--0.12    &    spec       &   den contrast & LS    &       $\geq$8.70      & none \\
\cite{Mosleh2018}           & 0.01--0.05    & spec      & Cls/F        & LS    & $\geq$9.70  & none \\
\cite{Tran17}               & $\sim 2.00$   & spec      & Cls/F        & LS    & $\geq$8.70  & none \\
\cite{Shankar2014}          & 0.05--0.20    & spec      & Cls/F        & LS    & $\geq$10.90 & none \\
\cite{Newman2014}           & $\sim 1.80$   & spec      & Cls/F        & LS    & $\geq$10.20 & none \\
\cite{Huertas13a}           & $<0.09$       & spec      & Cls/F        & LS & $\geq$10.20 & none \\

\cite{Strazzullo2023}       & 1.40--1.70    & phot/spec & Cls/F        & LS    & $\geq$10.50 & none \\
\cite{Morishita2017}        & 0.20--0.70    &     phot/spec      & Cls/F        & LS    &         $\geq$7.50  & none \\
\cite{Saracco2017}          & 1.20--1.40    & phot/spec & Cls/F        & LS    & $\geq$9.20  & none \\
\cite{Kelkar2015}           & 0.40--0.80    & phot/spec & Cls/F        & LS    & $\geq$9.90  & none \\
\cite{Rettura2010}          & $\sim 1.20$   &   phot/spec        & Cls/F        & LS    &   $\geq$10.40          & none \\
\cite{Huertas-Company2013a} & 0.20--1.00    & phot/spec & Cls/F        & LS    & $\geq$10.20 & none \\
\cite{Maltby2010}           & $\sim 0.17$   & phot      & Cls/F        & LS    & $\geq$9.70  & none \\
This work                   & 0.01--0.125   & spec      & $f(R_p)$ \& $f(M_h)$        & SS & $\geq$10.00 & none \\
\cite{Huertas13a}           & $<0.09$       & spec      &$f(R_p)$ \& $f(M_h)$        & SS & $\geq$10.20 & none \\
\hline
\hline
\multicolumn{7}{c}{Environmental Dependence} \\
\hline
\cite{Perez2025}            & 0.01--0.05    & spec      & Cls/F        & LS    & $\geq$8.20 & $R_\mathrm{{CLs}}>R_\mathrm{F}$ \\
\cite{Afanasiev2023}        & 1.40--2.80    & spec      & Cls/F        & LS    & $\geq$10.20 & $R_\mathrm{{CLs}}>R_\mathrm{F}$ \\
\cite{Noordeh21}            & 1.90--2.10    &  spec         & Cls/F        & LS    &       $\geq$10.00      & $R_\mathrm{{CLs}}>R_\mathrm{F}$ \\
\cite{Yoon2017}             & 0.01--0.15    &      spec     & den contrast & LS    &     $\geq$10.70        &  $R_\mathrm{{CLs}}>R_\mathrm{F}$ \\
\cite{Delaye2014}           & 0.80--1.50    &     spec      & Cls/F        & LS    &     $\geq$10.20        &  $R_\mathrm{{CLs}}>R_\mathrm{F}$  \\
\cite{Cooper2012}           & 0.40--1.20    &   spec        & den contrast & LS    &     $\geq$10.00        &  $R_\mathrm{{CLs}}>R_\mathrm{F}$ \\

\cite{Strazzullo13}         & $\sim 2.00$   &    phot/spec       & Cls/F        & LS    &      $\geq$10.00        &   $R_\mathrm{{CLs}}>R_\mathrm{F}$ \\
\cite{Ghosh24}              & 0.30--0.70    & phot      & den contrast & LS    & $\geq$8.60  & $R_\mathrm{{CLs}}>R_\mathrm{F}$ \\
\cite{Lani2013}             & 0.50--2.00    &   phot        & den contrast & LS    &          $\geq$10.00   &  $R_\mathrm{{CLs}}>R_\mathrm{F}$ \\
\cite{Papovich2012}         & 1.5           &    phot       & Cls/F        & LS    &      $\geq$10.20        &  $R_\mathrm{{CLs}}>R_\mathrm{F}$ \\

\cite{Chamba24}             & $<0.02$       & spec      & Cls/F        & LS    & $\geq$4.70 & $R_\mathrm{{CLs}}<R_\mathrm{F}$ \\
\cite{Xu2023}               & 2.49--2.52    &  spec         & Cls/F        & LS    & $\geq$10.20 & $R_\mathrm{{CLs}}<R_\mathrm{F}$ \\
\cite{Matharu2021}          & 0.86--1.34    &    spec       & Cls/F        & LS    &        $\geq$9.30     & $R_\mathrm{{CLs}}<R_\mathrm{F}$ \\
\cite{Demers2019}           & 0.01--0.06    & spec          & Cls/F        & LS    &    9.50-9.90     & $R_\mathrm{{CLs}}<R_\mathrm{F}$ \\
\cite{Matharu19}            & 0.86--1.34    &   spec        & Cls/F        & LS    &    $\geq$9.60 &   $R_\mathrm{{CLs}}<R_\mathrm{F}$      \\
\cite{Cebrian2014}          & 0.01--0.12    &    spec       & den contrast & LS    &        $\geq$10.00     & $R_\mathrm{{CLs}}<R_\mathrm{F}$ \\
\hline
\end{tabular}
\begin{tablenotes}\item[$\ast$] Here, LS denotes large-scale environment, corresponding to comparisons between broad environments such as clusters and the field or between regions of different overall galaxy density, while SS denotes small-scale environment, corresponding to local or intra-halo variations such as cluster-centric radius or host halo mass. Cls/F indicates studies comparing cluster and field galaxies, whereas den contrast refers to studies comparing galaxies in high-density and low-density environments. The ``Trend'' column indicates the reported direction of the environmental dependence when a significant effect is found.
\end{tablenotes}
\end{table*}

\section{An Open Question: Environmental Effects on the Galaxy Size--Mass Relation} \label{sec:comparison}
Over the past decades, numerous studies have investigated the impact of environment on the galaxy SMR, yet their findings remain inconsistent. While some report a significant dependence on environment, others find little to no effect. 
To better understand the source of these discrepancies, we compile the findings of more than 30 studies that examine the environmental dependence of the SMR.  Table~\ref{tab:env} summarizes the key characteristics of these works, including the redshift range, redshift quality, environmental definition, physical scale probed, stellar-mass range, and the reported trend. 
To facilitate comparison across the literature, Table~\ref{tab:env} distinguishes between large-scale and small-scale environment measures and identifies whether each study is based on cluster versus field comparisons or on contrasts between high- and low-density regions. This framework enables a structured analysis of the key methodological factors that may underlie the discrepancies in the literature, as discussed below.
\begin{itemize}[leftmargin=*, label=\textbullet]

\item Environmental definition:
The literature adopts a range of environmental tracers, including cluster versus field comparisons and statistical density contrasts. Table~\ref{tab:env} shows that conflicting conclusions arise even among studies using similar environmental definitions, with both cluster-based and density-based approaches yielding a mixture of null and positive results. To test this directly, we applied a statistical environment classification to our dataset and again found no evidence of an environmental effect. Although these results are not shown here, they are fully consistent with our main conclusions. This suggests that the adopted environmental tracer alone cannot account for the discrepancies in the literature.
\item Redshift quality:
The studies compiled in Table~\ref{tab:env} include spectroscopic, photometric, and mixed datasets. Both spectroscopic and photometric or mixed studies report a combination of environmental dependence and environmental independence. This indicates that redshift quality, although important for measurement reliability, is not by itself sufficient to account for the differing conclusions.
\item Stellar mass measurement and mass range:
The studies in Table~\ref{tab:env} span a broad range of stellar-mass limits and employ different methods to estimate stellar mass. No clear trend emerges linking the reported environmental dependence to the adopted stellar-mass method or mass threshold alone. To test this directly, we repeated our analysis using photometrically derived stellar masses from the Granada group \footnote{\url{https://www.sdss4.org/dr17/spectro/galaxy_granada/}}, based on the publicly available Flexible Stellar Population Synthesis code \citep{Conroy09}, instead of our default spectroscopic estimates. We again found no evidence of an environmental effect. The results, omitted here for brevity, align with our main findings.
\item Galaxy size definition: The choice of size definition does not account for the discrepancies. While some studies use effective (half-light) radius, others rely on metrics such as the semi-major axis or Petrosian radius. Nevertheless, both positive and null environmental effects are reported across all methods, indicating that the size definition is not the dominant factor.
\item Redshift range: We also investigated whether the redshift range of galaxy samples explains the conflicting results. Some studies focused on low-redshift galaxies ($z \lesssim 0.3$), while others examined populations at higher redshifts ($z \sim 1$–2). However, environmental effects are reported both at low and high redshifts, with no clear trend emerging. This suggests that redshift alone is unlikely to be the primary driver of the discrepancies.
\end{itemize}

An additional source of inconsistency is that even among studies reporting a significant environmental dependence, the direction of the effect is not uniform. As shown in Table~\ref{tab:env}, some studies find that cluster galaxies are larger than field galaxies at fixed stellar mass, whereas others report that field galaxies are larger than their cluster counterparts. This indicates that the disagreement in the literature is not limited to whether an environmental effect exists, but also extends to the sign of the reported effect.

While each of the above factors may influence the outcome to some extent, none of them individually accounts for the variation in reported environmental effects. Instead, the discrepancies likely arise from a combination of factors, including how stellar masses and galaxy sizes are measured, whether redshifts are photometric or spectroscopic, how galaxies are classified, the definition of environment, the redshift range probed, and the number of galaxies included in the sample. In particular, studies with small samples may suffer from limited statistical power, while larger samples often rely on photometric redshifts and mass estimates, which introduce their own uncertainties. The interplay among these elements makes it difficult to isolate a dominant source of disagreement. This suggests that conclusions about the presence or absence of environmental effects on the size--mass relation must be interpreted within the broader context of each study’s design, assumptions, and data quality.

Several aspects of our methodology give us confidence in the reliability of our findings. Our analysis is built on a large and statistically powerful spectroscopic sample drawn from the SDSS, allowing for precise measurements and robust conclusions. To ensure a clear environmental contrast, we adopt extreme definitions for both cluster and field galaxies: cluster members are selected using the GalWCat19 spectroscopic catalog, while field galaxies are drawn from the lowest-density regions, well outside any known structures. We exclude sparse data from the main sample that could bias the results due to low-number statistics. For structural measurements, we rely on a recent, high-quality galaxy size catalog with improved measurement efficiency and consistency. Stellar masses are taken from the MPA/JHU catalog, which derives them directly from galaxy spectra, ensuring reliability. Together, these methodological choices minimize uncertainties and observational biases, giving us high confidence in the robustness of our result: that the size--mass relation does not show a significant dependence on environment when controlling for galaxy type and stellar mass.

\section{Conclusion}\label{sec:conc}

In this study, we investigated the environmental dependence of the galaxy size–stellar mass relation (SMR) at low redshift ($z \leq 0.125$) using a large spectroscopic sample drawn from the SDSS-DR13 survey. By leveraging the GalWCat19 galaxy cluster catalog and a reliable void classification method, we were able to probe the SMR across a range of environments and galaxy populations.

\begin{itemize}[leftmargin=*, label=\textbullet]

\item Our analysis explored three layers of environmental influence. First, we compared galaxies in high-density (cluster) and low-density (void) environments. Second, we examined the SMR as a function of cluster mass by dividing clusters into low-, intermediate-, and high-mass subsamples. Third, we studied galaxies located at different positions within clusters—core, virial, and infall regions. To control for intrinsic differences among galaxies, we applied three independent classification schemes based on specific star formation rate, optical color, and bulge-to-total (B/T) light ratio.

\item Across all environments and classification methods, we find no significant variation in the SMR at fixed stellar mass and galaxy type. The size–mass trends for cluster and void galaxies are nearly identical, and no measurable differences are observed across cluster mass bins or cluster-centric regions. These results indicate that the structural scaling relations of galaxies are largely insensitive to their large-scale environment when intrinsic properties are accounted for.

\item We also observe that early-type galaxies (quiescent, red, or bulge-dominated) have steeper SMR slopes than late types (star-forming, blue, or disk-dominated), consistent with their more compact structures and distinct evolutionary histories. Furthermore, the size differences among galaxy types diminish at high stellar masses ($\log M_\ast \gtrsim 11$), suggesting that massive galaxies converge in size regardless of their morphology or star formation activity.

\item Overall, our findings demonstrate that galaxy size in the local universe is primarily driven by internal processes such as star formation, feedback, and mergers, rather than by external environmental influences. This challenges the expectations from hierarchical models that predict stronger environmental dependence, especially for early-type systems. Future work extending to higher redshifts and incorporating hydrodynamical simulations may help further clarify the interplay between internal and external drivers of galaxy evolution.
\end{itemize}

\section*{Acknowledgments}
We thank the reviewer for their thoughtful comments and suggestions, which improved the quality of this paper.
GW gratefully acknowledges support from the National Science Foundation through grant AST-2205189.

\appendix
\setcounter{section}{0}
\setcounter{figure}{0}
\setcounter{table}{0}
\setcounter{equation}{0}

\renewcommand{\thesection}{Appendix~\Alph{section}}
\renewcommand{\thefigure}{\Alph{section}\arabic{figure}}
\renewcommand{\thetable}{\Alph{section}\arabic{table}}

\makeatletter
\@addtoreset{figure}{section}
\@addtoreset{table}{section}
\makeatother
\section{Systematic Effects of Void Selection Criteria}
\label{appendix:voids}
\begin{figure*}
    \centering\hspace{-0.25cm}
    \includegraphics[width=0.8\linewidth]{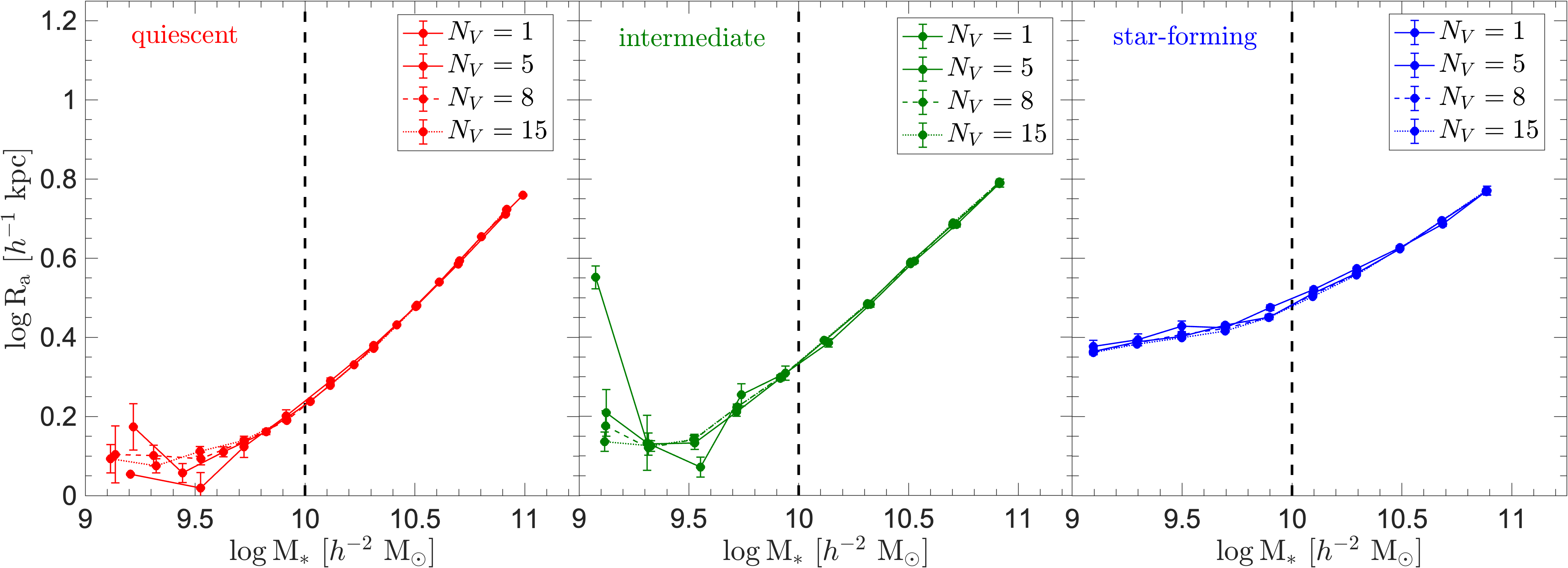} 
    \caption{Comparison of the size-stellar mass relation (SMR) for quiescent (red, left panel), intermediate (green, middle panel), and star-forming (blue, right panel) galaxies, using four different void definitions containing 1 (solid lines), 5 (dashed lines), 8 (dotted lines), or 15 (dot--dashed lines) galaxies. The overlap of the lines in each panel indicates that the SMR relation remains consistent across all void definitions, with the lines being nearly indistinguishable.
    }
    \label{fig:similarinvoids}
\end{figure*}

To assess the impact of our void galaxy definition on SMR, we examine how the results vary when changing the number of galaxies used to define the low-density environment. In our main analysis, a galaxy is identified as a void galaxy if it lies at the center of a cylinder containing exactly eight galaxies (including itself), within a radius of $2~h^{-1}$~Mpc and a velocity depth of $3000~\mathrm{km~s^{-1}}$. 
To test the sensitivity of this criterion, we repeat the analysis using alternative definitions: cylinders containing 1, 5, and 15 galaxies, in addition to the baseline value of 8. 
Figure~\ref{fig:similarinvoids} compares the SMR for galaxies identified under these four void definitions. 
As shown in Figure~\ref{fig:similarinvoids}, the SMR relation remains consistent across the different void thresholds. This demonstrates that our results are not affected by the specific choice of void selection criteria and confirms the robustness of our conclusions regarding galaxies in low-density environments.  We also tested an alternative low-density selection based on a nearest-neighbor metric and found consistent SMR trends, further supporting the robustness of our void-sample definition.

\section{Size-Mass Relation Using Color and Morphology Across Cluster Mass and Radius} \label{appendix:color_bt}
To verify the robustness of our results presented in Section \ref{sec:galclass}, we repeat the size–stellar mass relation (SMR) analysis using alternative galaxy classification types based on optical color and B/T ratio. In both cases, galaxies are divided into three types: red, green, and blue based on color, and bulge-dominated, intermediate, and disk-dominated based on B/T ratio, as described in Section~\ref{sec:galclass}.

\begin{figure*}
    \centering
    \includegraphics[width=0.8\linewidth]{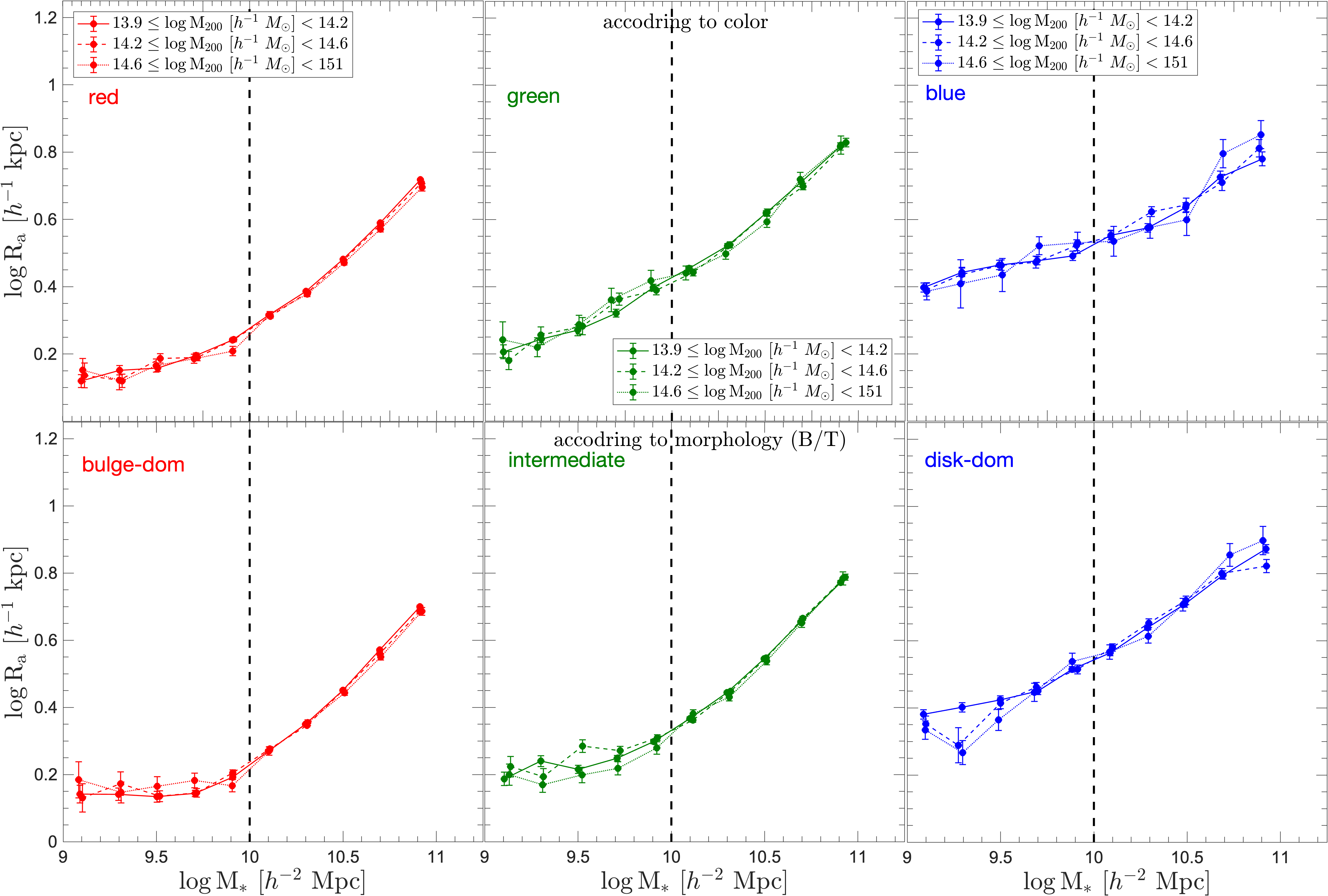} 
    \vspace{-0.1cm}
    \caption{Size-stellar mass relation (SMR) for galaxies in $\mathtt{GalWCat19}$ clusters, divided into three mass subsamples: $13.9 \leq\log\mathrm{M_{200}} < 14.2~[h^{-1}\mathrm{M_{\odot}}]$ (solid lines), $14.2 \leq \log\mathrm{M_{200}} < 14.6~[h^{-1}~\mathrm{M_{\odot}}]$ (dashed lines), and $14.6 \leq \log\mathrm{M_{200}} \leq 15.1~[h^{-1}~\mathrm{M_{\odot}}]$ (dotted lines). Upper panels: galaxies are divided into red, green, and blue based on color. Lower panels: galaxies are divided into bulge-dominated, intermediate. and disk-dominated based on B/T ratio.}
    \label{fig:masscolorBT}
\end{figure*}

Figure~\ref{fig:masscolorBT} presents the SMR for galaxies in different cluster mass bins. The upper panels show the SMR for color-based classifications, and the lower panels show the SMR for B/T-based classifications. In each case, galaxies are further separated into low-mass (solid lines), intermediate-mass (dashed lines), and high-mass (dotted lines) clusters. In the upper panels, galaxies are classified as red, green, or blue according to their location in the color–magnitude diagram, as described in Section~\ref{sec:galclass}. The lower panels follow the same format but use structural classification based on the B/T ratio, separating galaxies into bulge-dominated, intermediate, and disk-dominated systems, as described in Section~\ref{sec:galclass}.
The results are consistent with those obtained using the SFR-based classification (see Section \ref{sec:mass}). For each galaxy type, the SMR trends remain nearly identical across all cluster mass bins, confirming that the observed independence of the SMR from cluster mass does not depend on the choice of classification type.

\begin{figure*}
    \centering
 \includegraphics[width=0.8\linewidth]{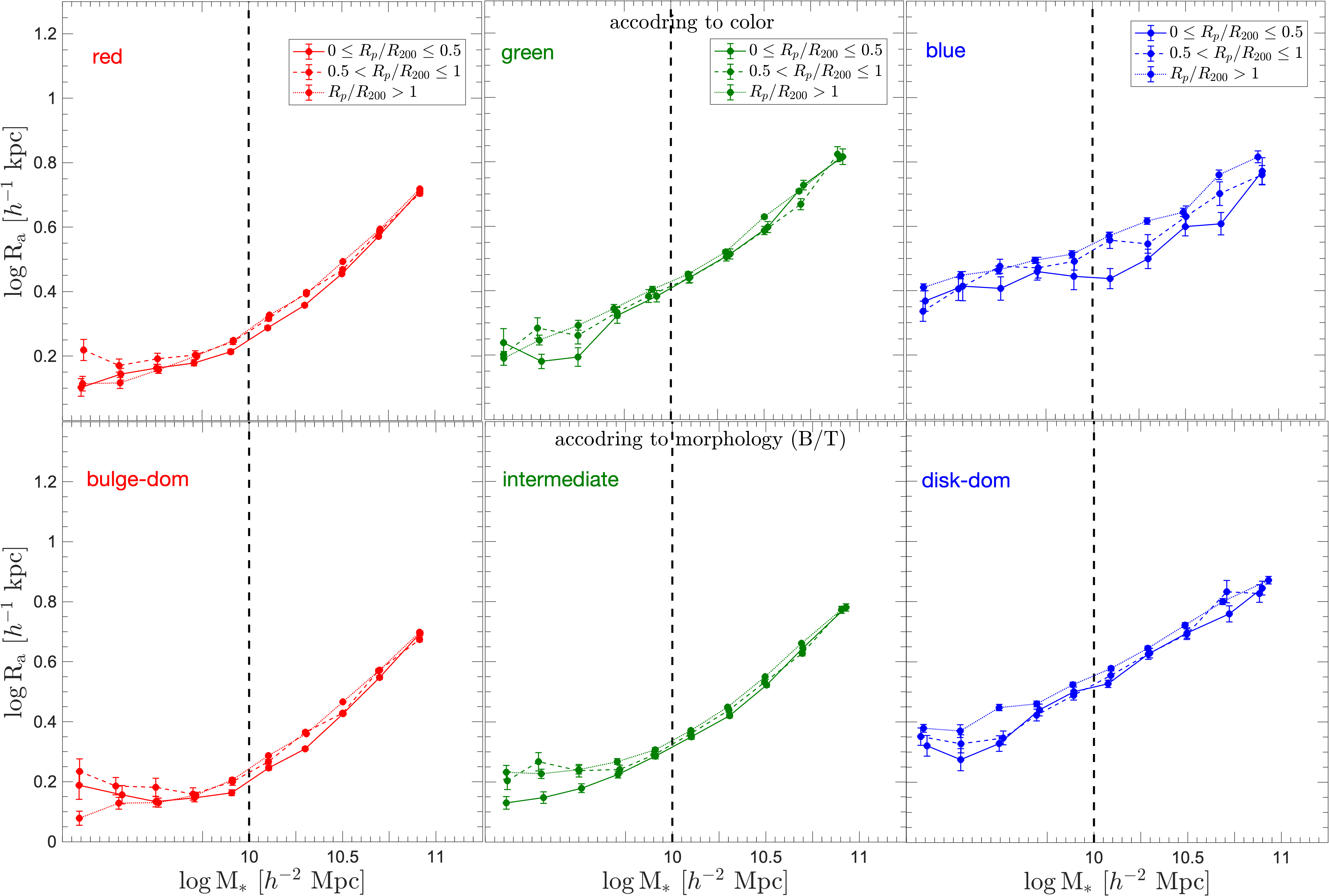} 
    \vspace{-0.1cm}
    \caption{Size-stellar mass relation (SMR) for galaxies in $\mathtt{GalWCat19}$ clusters, divided into three cluster-centric radial zones subsamples: \( 0 \leq \mathrm{R_p}/R_{200}< 0.5\) (solid lines), \( 0.5 \, \leq \mathrm{R_p} /R_{200}< 1 \) (dashed lines), and \(\mathrm{R_p}/R_{200} \geq 1 \) (dotted lines). Upper panels: galaxies are divided into red, green, and blue based on color. Lower panels: galaxies are divided into bulge-dominated, intermediate and disk-dominated based on B/T ratio.}
    \label{fig:radiuscolorBT}
\end{figure*}

Figure~\ref{fig:radiuscolorBT} presents the SMR for galaxies in different cluster regions. The upper panels show the SMR for color-based classifications, and the lower panels show the SMR for B/T-based classifications, with the same notations as Figure~\ref{fig:masscolorBT}. The results are consistent with those obtained using SFR-based classification (see Section~\ref{sec:radius}). The SMR shows no significant dependence on cluster region across all classification methods. 

\bibliography{Ref}{}
\bibliographystyle{aasjournal}
\end{document}